\documentclass[aip,graphicx]{revtex4-1}
\usepackage{graphicx}
\DeclareGraphicsExtensions{.pdf,.jpg,.png,.eps}
\usepackage{amsfonts}
\usepackage{amsmath}
\usepackage{amssymb}
\usepackage{color}
\usepackage{color}
\usepackage{epstopdf}

\begin{document}
\title{Magnetic structures and magnetic phase transitions in the Mn-doped orthoferrite TbFeO$_3$ studied by neutron powder diffraction}
%
%
%
\author{Harikrishnan S. Nair}
\email{h.nair.kris@gmail.com, hsnair@uj.ac.za}
\affiliation{Highly Correlated Matter Research Group, Physics Department, P. O. Box 524, University of Johannesburg, Auckland Park 2006, South Africa}
\author{Tapan Chatterji}
\affiliation{Institut Laue-Langevin, BP 156, 38042 Grenoble Cedex 9, France}
\author{C. M. N. Kumar}
\affiliation{J\"{u}lich Centre for Neutron Science JCNS, Outstation at SNS, Oak Ridge National Laboratory, Oak Ridge, TN 37831, USA}
\affiliation{Chemical and Engineering Materials Division, Oak Ridge National Laboratory, Oak Ridge, TN 37831, USA}
\author{Thomas Hansen}
\affiliation{Institut Laue-Langevin, BP 156, 38042 Grenoble Cedex 9, France}
\author{Hariharan Nhalil}
\affiliation{Department of Physics, Indian Institute of Science, Bangalore 560012, India}
\author{Suja Elizabeth}
\affiliation{Department of Physics, Indian Institute of Science, Bangalore 560012, India}
\author{Andr\'{e} M. Strydom}
\affiliation{Highly Correlated Matter Research Group, Physics Department, P. O. Box 524, University of Johannesburg, Auckland Park 2006, South Africa}
\affiliation{Max Planck Institute for Chemical Physics of Solids (MPICPfS), N\"{o}thnitzerstra{\ss}e 40, 01187 Dresden, Germany}
\date{\today}
\begin{abstract}
The magnetic structures and the magnetic phase transitions in the Mn-doped orthoferrite TbFeO$_3$ studied using neutron powder diffraction are reported. Magnetic phase transitions are identified at $T^\mathrm{Fe/Mn}_N \approx$ 295~K where a paramagnetic-to-antiferromagnetic transition occurs in the Fe/Mn sublattice, $T^\mathrm{Fe/Mn}_{SR} \approx$ 26~K where a spin-reorientation transition occurs in the Fe/Mn sublattice and $T^\mathrm{R}_N \approx$ 2~K where Tb-ordering starts to manifest. At 295~K, the magnetic structure of the Fe/Mn sublattice in TbFe$_{0.5}$Mn$_{0.5}$O$_3$ belongs to the irreducible representation $\Gamma_4$ ($G_xA_yF_z$ or $Pb'n'm$). A mixed-domain structure of ($\Gamma_1 + \Gamma_4$) is found at 250~K which remains stable down to the spin re-orientation transition at $T^\mathrm{Fe/Mn}_{SR}\approx$ 26~K. Below 26~K and above 250~K, the majority phase ($> 80\%$) is that of $\Gamma_4$. Below 10~K the high-temperature phase $\Gamma_4$ remains stable till 2~K. At 2~K, Tb develops a magnetic moment value of 0.6(2)~$\mu_\mathrm{B}/$f.u. and orders long-range in $F_z$ compatible with the $\Gamma_4$ representation. Our study confirms the magnetic phase transitions reported already in a single crystal of TbFe$_{0.5}$Mn$_{0.5}$O$_3$ and, in addition, reveals the presence of mixed magnetic domains. The ratio of these magnetic domains as a function of temperature is estimated from Rietveld refinement of neutron diffraction data. Indications of short-range magnetic correlations are present in the low-$Q$ region of the neutron diffraction patterns at $T < T^\mathrm{Fe/Mn}_{SR}$. These results should motivate further experimental work devoted to measure electric polarization and magnetocapacitance of TbFe$_{0.5}$Mn$_{0.5}$O$_3$.
\end{abstract}
%
\maketitle
\section{\label{INTRO}Introduction}
\indent 
The orthoferrite [$R$FeO$_3$; $R$ = rare earth] 
oxides have been recently re-investigated experimentally and 
theoretically from the fascinating perspective of multiferroicity.
\cite{mandal2011spin,shang2013multiferroic,deng2015magnetic,pavlov2012optical,zhao2014creating}
Pursuing the recent line of multiferroics research, 
theoretical work on $R$FeO$_3$ thinfilms has identified 
that strain can convert paraelectric phase of 
orthoferrites in to ferroelectrics thus rendering 
them multiferroic\cite{zhao2014creating}. It has 
been found theoretically that for large values 
of strain on $R$FeO$_3$ with large rare earth ion, 
giant polarization is realized. In fact, with increasing
strain, a new ferroelectric phase, 
not observed in any perovskite before, 
is realized. Multifunctional properties like 
large magnetoelectric coupling and ultrafast optical 
control of spins have been observed in the orthoferrites
\cite{tokunaga2008magnetic,yamaguchi2013terahertz,mikhaylovskiy2014terahertz}.
The $R$FeO$_3$ realize high N\'{e}el temperature, 
$T_N \approx$ 623 -740~K\cite{marezio1970crystal,eibschutz1967mossbauer}
however, in bulk form they are paraelectric 
rather than ferroelectric suggesting weak 
multiferroic effects. Weak ferroelectric polarization
has been recently reported in Gd and Sm orthoferrites
\cite{tokunaga2008magnetic,lee2011spin} which are thought to 
have "improper" origin induced by magnetic order. 
In TbFeO$_3$, an unusual incommensurate magnetic 
phase was discovered through neutron diffraction
\cite{artyukhin2012solitonic}
-- it was shown that the exchange of 
spin waves between extended topological defects could result in 
novel magnetic phases drawing parallels with the Yukawa
forces that mediate between protons and neutrons in a nucleus. 
The Fe$^{3+}$ moments in TbFeO$_3$ exhibit 
$G_xA_yF_z$ ($Pb'n'm$) spin configuration at room temperature
\cite{bouree1975mise,bertaut1967structures,tejada1995quantum} which is
accompanied by a spin-reorientation to $F_xC_yG_z$ ($Pbn'm'$). 
At 3~K, another spin re-orientation occurs to revert to the
$G_xA_yF_z$ ($Pb'n'm$) structure. It is considered that the Tb$^{3+}$ 
spins order in $F_xC_y$ structure 
in 10 - 3~K interval and in the $A_xG_y$ structure below 3~K. 
Doping the $R$-site in $R$FeO$_3$ with another rare 
earth is found to be profitable to realize electric field induced
generation and reversal of ferromagnetic moments
\cite{tokunaga2012electric,tokunaga2014magnetic}. 
\begin{figure}[!t]
\centering
\includegraphics[ scale=0.42]{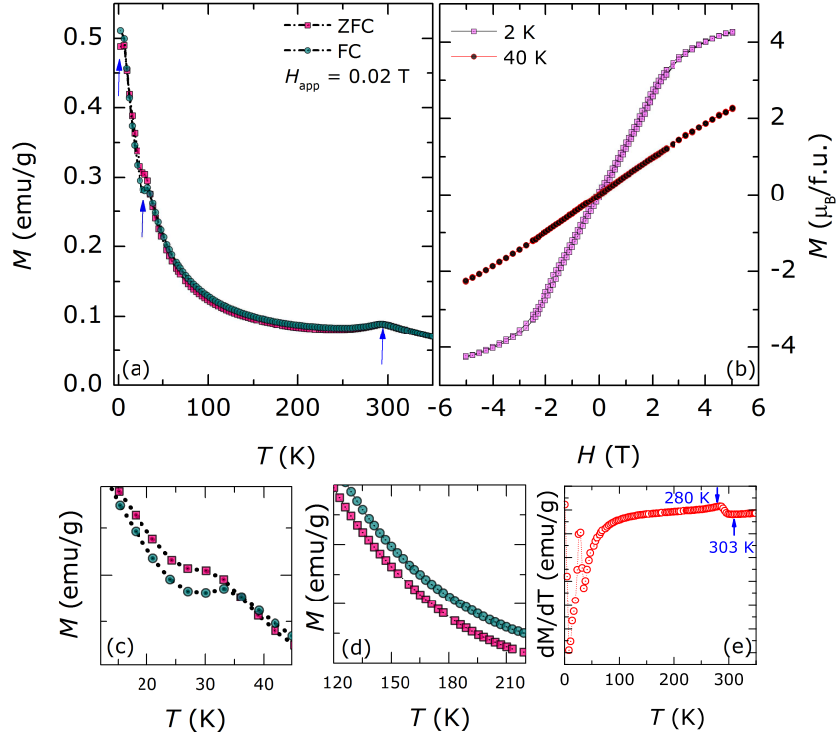}
\caption{\label{fig_mag} (color online) (a) The magnetization data of TbFe$_{0.5}$Mn$_{0.5}$O$_3$ in ZFC and FC protocols at applied field of 0.02~T. Three transitions are visible at $T^\mathrm{Fe/Mn}_{N}\approx$ 295~K, $T^\mathrm{Fe/Mn}_{SR}\approx$ 26~K and at $T^\mathrm{R}_{N}\approx$ 2~K. (b) The isothermal magnetization TbFe$_{0.5}$Mn$_{0.5}$O$_3$ at 2~K and at 40~K. (c) Magnifies the transition at $T^\mathrm{Fe/Mn}_{SR}$ while (d) shows the irreversibility present till $T^\mathrm{Fe/Mn}_{N}$. The derivative $dM/dT$ is shown in (e) to clarify the breadth of the transition at $T^\mathrm{Fe/Mn}_{N}$.}
\end{figure}
Chemical substitution at the Fe-site in $R$FeO$_3$ also 
brings about interesting multiferroic effects.
For example, in the case of Mn-substituted 
YFeO$_3$, magnetoelectric 
and magnetodielectric effects at different temperatures
were reported\cite{mandal2011spin}. First-order spin-reorientation 
effects were observed as a result of Mn-substitution
however, the magnetodielectric effects were observed at lower 
temperatures than $T_N$ or $T_{SR}$.
Giant magnetodielectric coupling is also observed in another 
doped-orthoferrite, DyMn$_{0.33}$Fe$_{0.67}$O$_3$
\cite{hong2012temperature}.
Spin-reorientation effects and magnetic sublattice effects 
were also observed in doped-orthoferrites
with large $R$\cite{nagata2001magnetic,mihalik2013magnetic}. 
$G$-type magnetic ordering of Mn$^{3+}$ and 
Cr$^{3+}$ spins were observed below $T_N \approx$ 84~K in the
case of TbMn$_{0.5}$Cr$_{0.5}$O$_3$\cite{staruch2014magnetic}, 
in addition to signatures of short-range magnetic correlations
observed below 40~K which was attributed to the ferromagnetic 
component from canting of magnetic moments along the $c$-axis.
In the case of Mn-substituted compound TbFe$_{0.75}$Mn$_{0.25}$O$_3$, 
the $T_N$ was determined to be 550~K and the $T_{SR}$ as 180~K through 
magnetic studies and M\"{o\ss}bauer spectroscopy\cite{kim2011spin}.
\\
\begin{figure}[!b]
\centering
\includegraphics[ scale=0.62]{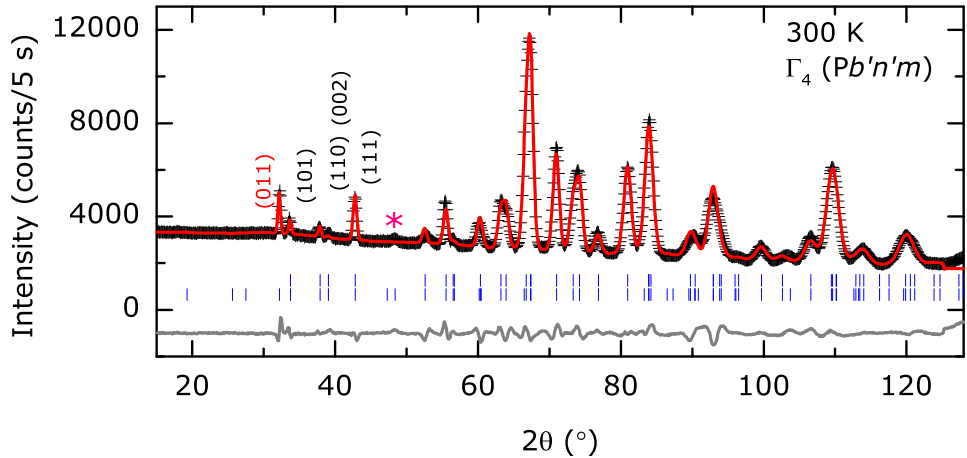}
\caption{\label{fig_npd_300} (color online) The Rietveld refinements of the neutron powder diffraction pattern of TbFe$_{0.5}$Mn$_{0.5}$O$_3$ at 300~K. The nuclear structure is refined using the space group $Pbnm$. The magnetic structure conforms to the representation $\Gamma_4$ or $G_xA_yF_z$. The lower set of vertical marks in the figure correspond to the magnetic unitcell and the upper ones to the nuclear. Some of the reflections in the low-2$\Theta$ region are shown indexed. An unknown impurity peak is marked with an asterisk.}
\end{figure}
\indent
In our previous investigation using magnetometry it was inferred that 
TbFe$_{0.5}$Mn$_{0.5}$O$_3$ orders in $A_xG_yC_z$ ($\Gamma_1$) 
structure at $T^\mathrm{Fe/Mn}_N \approx$ 286~K followed 
by a spin re-orientation at $T^\mathrm{Fe/Mn}_{SR} 
\approx$ 28~K to the structure 
$G_xA_yF_z$ ($\Gamma_4$)\cite{hariharan2015reorientation}. 
No signature of Tb ordering was obtained in the previous study. 
In the present manuscript, we make 
a detailed investigation of the magnetic structures and spin 
re-orientation transitions in TbFe$_{0.5}$Mn$_{0.5}$O$_3$ 
using neutron powder diffraction in order to confirm the magnetic 
structure arrived at through macroscopic magnetization
earlier. We update the magnetic structures as a function of 
temperature and observe that they evolve between $\Gamma_1$
and $\Gamma_4$ through mixed-domains of ($\Gamma_1$ + $\Gamma_4$).
\section{\label{EXP}Experimental details}
\indent
Polycrystalline samples of TbFe$_{0.5}$Mn$_{0.5}$O$_3$ 
were prepared by conventional solid state 
reaction methods employing the oxides 
Tb$_2$O$_3$, FeO, MnO$_2$ 
(all from Sigma Aldrich, 99.9$\%$) 
as precursors. The thoroughly-mixed powder was 
heated at 1300$^{\circ}$~C for 4 days
with intermediate grinding. The phase-purity 
of the black powder that resulted was checked 
using x ray diffraction employing a Philips X'pert diffractometer
with Cu-$K\alpha$ radiation.
The chemical composition of the
prepared sample was determined using the
Inductively Coupled Plasma emission Spectroscopy
(ICPAES) method. 
Magnetization measurements were 
performed on a sintered pellet of the 
sample in a magnetic property measurement
system (MPMS, Quantum Design, San Diego). 
Neutron powder diffraction experiments 
were performed on 8~g of 
TbFe$_{0.5}$Mn$_{0.5}$O$_3$ powder
at the instrument D1B in ILL, Grenoble. 
A wavelength of 2.52~\AA~was used for the 
experiment. Diffractograms were recorded at 2~K, 5~K, 10~K, 26~K, 
50~K to 100~K in 10~K interval and 100~K to 300~K in 50~K interval. 
The diffraction data was analyzed using
FullProf suite of programs\cite{fullprof} employing the 
Rietveld method\cite{rietveld}. 
Magnetic structure was determined using the software 
SARA$h$\cite{sarah_wills} and was refined using FullProf.
\begin{figure}[!b]
\centering
\includegraphics[ scale=0.45]{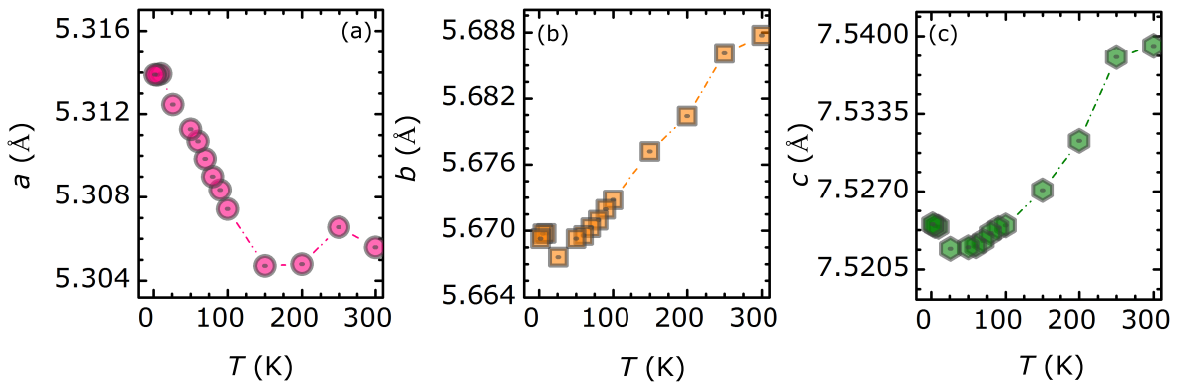}
\caption{\label{fig_latt} (color online) The variation of the lattice parameters $a$, $b$ and $c$ of TbFe$_{0.5}$Mn$_{0.5}$O$_3$  as a function of temperature. $b$ and $c$ follow a similar temperature-dependence while $a$ shows an increase at low temperature.}
\end{figure}
\section{\label{RESULTS} Results}
\subsection{\label{mag} Magnetization}
\indent
The experimentally measured magnetization curves, $M(T)$, 
in zero field-cooled and field-cooled protocols at 0.02~T are 
plotted in Fig~\ref{fig_mag} (a) and the isothermal 
magnetization curves of TbFe$_{0.5}$Mn$_{0.5}$O$_3$ 
at 2~K and 40~K are presented in (b).  Three 
magnetic phase transitions are identified in (a) {\em viz.,}
$T^\mathrm{Fe/Mn}_{N} \approx$ 295~K, $T^\mathrm{Fe/Mn}_{SR} 
\approx$ 26~K and $T^\mathrm{R}_{N} \approx$ 2~K.
The magnetic phase transition at $T^\mathrm{Fe/Mn}_{N}$ marks 
the paramagnetic (PM) to antiferromagnetic (AFM) 
phase transition. TbFe$_{0.5}$Mn$_{0.5}$O$_3$ adopts 
$G_xA_yF_z$ magnetic structure below this temperature
\cite{hariharan2015reorientation}.
The apparent difference seen in the transition temperatures 
of the single crystal and the polycrystalline sample
might stem from differences in sample quality. 
As determined by the ICPAES method, the chemical
composition of the sample is Tb$_{1.97}$Fe$_{0.51}$Mn$_{0.49}$O$_3$
which is very close to the nominal value.
The second transition occurring 
at $T^\mathrm{Fe/Mn}_{SR} \approx$ 26~K 
corresponds to the spin-reorientation transition where the 
structure transforms from $G_xA_yF_z$ to $A_xG_yC_z$. 
At temperatures close to 2~K, signatures of Tb-order are observed 
which is reflected in the magnetization measurements at 0.02~T as 
a irreversibility between the ZFC and 
FC plots at $\approx$ 5~K. The panel (c) 
in Fig~\ref{fig_mag} magnifies the $M(T)$ curves around
$T^\mathrm{Fe/Mn}_{SR}$ where a "loop"-like feature is observed. 
It can be observed that the "loop"-like feature 
begins at $\approx$ 36~K and extends till about 18~K. 
The $M(T)$ at a higher applied field of 1~T
was measured where the signs of magnetic phase transitions 
or bifurcation between ZFC and FC were absent (data not shown).
The panel (d) shows the presence of irreversibility extends 
over a wide temperature range from 36~K to $T^\mathrm{Fe/Mn}_N$.
In the panel (e), the derivative $dM/dT$ versus $T$ is plotted
to show the $T^\mathrm{Fe/Mn}_{N}$ transition 
more clearly. It is seen that the
$T^\mathrm{Fe/Mn}_{N}$ is a very broad transition with a spread
in temperature from about 280~K extending to 303~K. The transition
temperature of 295~K is estimated approximately at the point of
steepest slope of $dM/dT$.
\\
\begin{table}[!b]
\caption{\label{tab1} The refined lattice parameters and fractional coordinates of TbFe$_{0.5}$Mn$_{0.5}$O$_3$ at 300~K, 150~K, 26~K and at 2~K. These parameters are obtained through Rietveld refinement of the neutron powder diffraction data obtained from the instrument D1B, ILL, Grenoble. The nuclear structure model used was $Pbnm$ with Fe/Mn at $4b$ ($\frac{1}{2}$, 0, 0) and Tb at $4c$ ($x$,$y$,$z$).}
\setlength{\tabcolsep}{8pt}
\begin{tabular}{llllll} \hline\hline
                      &   300~K    &     150~K    &   26~K      &    2~K                       \\ \hline\hline
$a (\AA)$   & 5.3055(5)  &    5.3046(5) &   5.3124(4) &    5.3138(1)      \\  
$b (\AA)$   & 5.6877(8)  &    5.6772(6) &   5.6676(8) &    5.6692(6)     \\
$c (\AA)$   & 7.5391(9)  &    7.5270(8) &   7.5221(2) &        7.5242(4)     \\
Tb: $x$              & -0.0173(7)     &   -0.0193(7)    &  -0.0223(6)    &  -0.0202(6)    \\
$y$                  & 0.0715(6)      &    0.0714(6)    &   0.0732(5)    &  0.0712(5)     \\
$z$                  & 0.25           &    0.25         &   0.25         &  0.25          \\
O1: $x$              & 0.1089(8)      &    0.1084(9)    &   0.1091(9)    &  0.1094(8)       \\
$y$                  & 0.4669(7)      &    0.4659(8)    &   0.4665(7)    &  0.4668(7)       \\
$z$                  & 0.25           &    0.25         &   0.25         &  0.25            \\
O2: $x$              & -0.2999(6)     &   -0.3007(6)    &  -0.3007(5)    & -0.3000(5)        \\
$y$                  & 0.3152(6)      &    0.3149(5)    &   0.3145(5)    &  0.31420(5)   \\
$z$                  & 0.0509(6)      &    0.0513(6)    &   0.0549(6)    &  0.0520(6)    \\ \hline\hline
\end{tabular}   
\end{table}
\begin{figure}[!t]
\centering
\includegraphics[ scale=0.60]{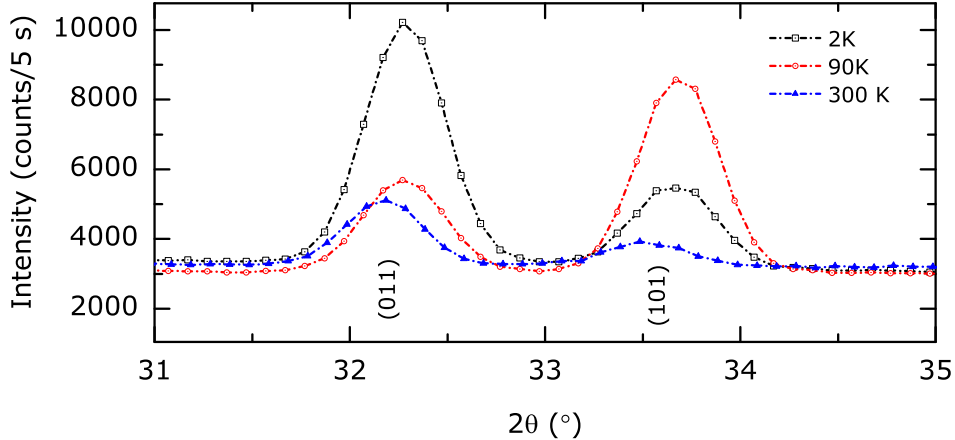}
\caption{\label{fig_npd} (color online) Expanded view of the neutron diffraction pattern of TbFe$_{0.5}$Mn$_{0.5}$O$_3$ at 300~K, 90~K and 2~K that shows the evolution of intensity of the (011) and the (101) reflections. $I_{(011)} > I_{(101)}$ is observed at 2~K while at 90~K, the opposite is true. At 300~K, again the condition $I_{(011)} > I_{(101)}$ is satisfied.}
\end{figure}
\indent
At 2~K, the maximum magnetization attained with application of 
5~T is about $4~\mu_\mathrm{B}/$f.u. 
This value is lower than the value obtained for ferromagnetic 
alignment of Fe$^{3+}$, Mn$^{3+}$ and
Tb$^{3+}$ moments. The observed maximum moment at 2~K, 5~T 
is comparable to the moment value obtained on the single crystal
of TbFe$_{0.5}$Mn$_{0.5}$O$_3$ in the case of $H \parallel b$ at 
25~K\cite{hariharan2015reorientation}. The first-order spin-flip-like 
transition observed at $H_c \pm$ 26~kOe is not 
clear in the present measurement on polycrystalline 
sample. Though a weak hysteresis is observed at 
2~K in Fig~\ref{fig_mag} (b), prominent hysteresis loops as observed 
for $H \parallel c$ in single crystals are absent. 
These features underline the magnetic anisotropy 
in the crystal sample of TbFe$_{0.5}$Mn$_{0.5}$O$_3$
\cite{hariharan2015reorientation}.
\subsection{\label{npd} Neutron diffraction}
\indent
The neutron diffraction patterns at different 
temperature points were refined using FullProf.
The values of refined lattice parameters and 
fractional atomic coordinates of 
TbFe$_{0.5}$Mn$_{0.5}$O$_3$ at 300~K, 150~K, 26~K and 2~K
are collected in Table~\ref{tab1}. The 
temperature-dependent variation of the 
lattice parameters $a$, $b$ and $c$ are shown in 
Fig~\ref{fig_latt}. No significant anomalies 
are observed for the unit cell parameters 
as a function of temperature
except for a change-of-slope at 150~K and 
250~K. However, lack of enough data points 
makes any inference unreliable here. 
The trend of thermal evolution 
of $b$ and $c$ are comparable while that of 
$a$ is opposite to the other two. 
Jahn-Teller (JT) or pseudo-JT distortion 
has been correlated with the observation 
of multiferroicity in perovskites\cite{bersuker2012pseudo} 
including TbMnO$_3$ which are known to show significant JT
effect\cite{zhou2007evidence}. In order to 
investigate the presence of JT effect in 
TbFe$_{0.5}$Mn$_{0.5}$O$_3$, estimates of the distortion
parameters and bond angles and bond 
distances were obtained from the refined 
structural data, Table~\ref{tab2}. 
However, with the substitution of 50$\%$ Fe 
at the Mn-site, the effects of
JT-distortion are found to have diminished in 
TbFe$_{0.5}$Mn$_{0.5}$O$_3$ however, effects 
of distortion of the perovskite
structure are clearly seen. To facilitate 
comparison, in Table~\ref{tab2}, the values 
of the JT-parameters of
TbMnO$_3$ at 300~K taken from Ref.[28] 
\cite{zhou2007evidence} 
are given in parenthesis.
\\
\begin{table*}[!t]
\caption{\label{tab2} The bond distances and bond angles of TbFe$_{0.5}$Mn$_{0.5}$O$_3$ at 300~K, 150~K, 26~K and at 2~K. These parameters are obtained through Rietveld refinement of the neutron powder diffraction data at the respective temperatures. The Jahn-Teller parameters are tabulated for different temperatures. For a comparison, the JT-values for TbMnO$_3$ at 300~K collected from Ref.[28] and given in parenthesis.}
\setlength{\tabcolsep}{4pt}
\begin{tabular}{llllll} \hline\hline
          &   300~K                  &   150~K               &        26~K               &    2~K                        \\ \hline\hline
Mn--O(2)  & $l$=2.123(4) $\times$ 2     &  $l$=2.112(5) $\times$2  &     $l$=2.114(6) $\times$2   &   $l$=2.111(5) $\times$2        \\  
          & $s$=1.947(6) $\times$ 2     &  $s$=1.949(6) $\times$2  &     $s$=1.956(7) $\times$2   &   $s$=1.951(7) $\times$2     \\  
Mn--O(1)  & $m$=1.977(8) $\times$ 2     &  $m$=1.977(7) $\times$ 2 &    $m$=1.977(8) $\times$2    &  $m$=1.978(6) $\times$2     \\
Mn--O(2)--Mn   & 145.81(8)              &    146.06(9)             &  145.14(7)                   &   145.81(11)                 \\
Mn--O(1)--Mn   & 144.9(12)               &    144.2(10)              &  144.04(8)                   &  144.9(9)                   \\
O(1)--Mn--O(1) & 180                 &  180                  &  180                      & 180                      \\
O(2)--Mn--O(2) & 90.57(12) $\times$ 2    &  90.60(8) $\times$ 2     &  90.16(11) $\times$ 2         &  90.41(8) $\times$ 2      \\
               & 89.53(9) $\times$ 2    &  89.40(6) $\times$ 2     &  89.83(9) $\times$ 2         & 89.59(7) $\times$ 2       \\ 
$Q_2 = 2(l -s)/\sqrt(2)$     & 0.2489 (0.45)  & 0.2305        &  0.2234    &  0.2263                                 \\
$Q_3 = 2(2m-l -s)/\sqrt(6)$  & -0.0947 (-0.2) &-0.0874        & -0.0947    & -0.0865                                  \\
$\phi$ = tan($Q_3/Q_2$)      & -20.84$^{\circ}$ (-24)  & -20.77$^{\circ}$  &  -22.98$^{\circ}$   &  -20.92$^{\circ}$   \\
$\rho_0$ = $\sqrt{Q_3^2 + Q_2^2}$ & 0.266 (0.5)        & 0.2465            & 0.2426              & 0.2423                \\  \hline\hline
\end{tabular}   
\end{table*}
\begin{figure}[!t]
\centering
\includegraphics[ scale=0.35]{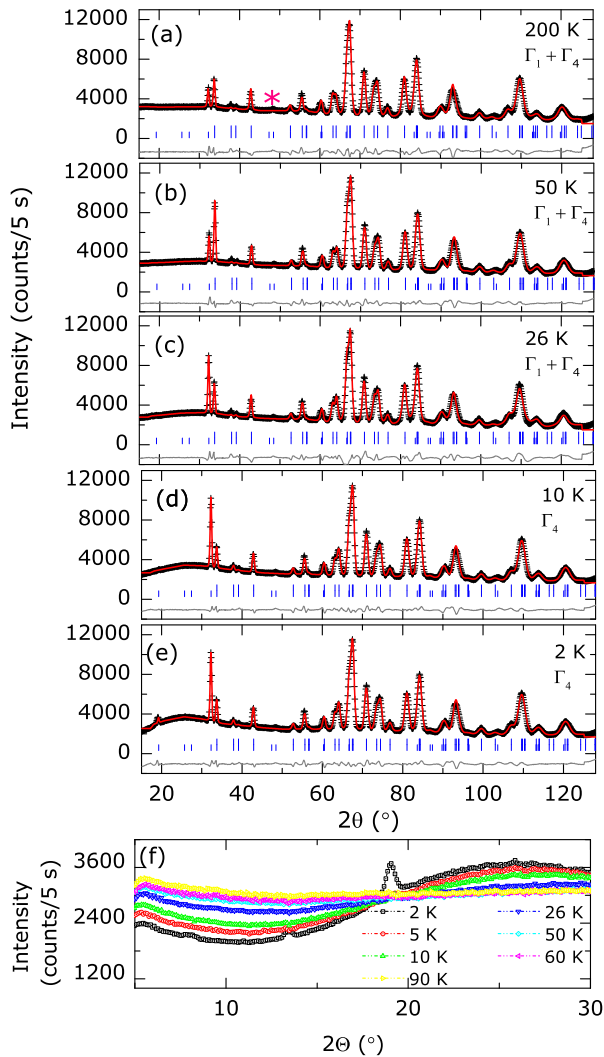}
\caption{\label{fig_npd_refine} (color online) The Rietveld refinements of the neutron diffraction pattern of TbFe$_{0.5}$Mn$_{0.5}$O$_3$ at (a) 200~K, (b) 50~K, (c) 26~K and (d) 10~K (e) 2~K. The different magnetic structures at each temperature is indicated in each panel. Recall that at 300~K (Fig. 2) $\Gamma_4$ prevails as a single phase. The unknown peak at $\approx$ 48~$^\circ$ is marked using an asterisk in (a). (f) Shows the presence of strong diffuse scattering in the low-angle region at $T < T^\mathrm{Fe/Mn}_{SR}$. Such a diffuse feature is related to the spin fluctuations in the Tb-sublattice.}
\end{figure}
\indent
A symmetry analysis of $R$FeO$_3$ in $Pbnm$ 
space group with Fe$^{3+}$ in $4b$ and $R^{3+}$ in $4c$
Wyckoff positions leads to eight irreducible 
representations, $\Gamma_1$ through $\Gamma_8$, 
for magnetic structure. For the $4b$ position, 
the configurations $\Gamma_5$ to $\Gamma_8$ 
are not allowed and hence
$\Gamma_1$, $\Gamma_2$, $\Gamma_3$ and $\Gamma_4$ 
are selected as the final possibilities. 
Table~\ref{tab3} lists these irreducible 
representations, the Shubnikov space groups 
and the magnetic structure notations used for 
$R$FeO$_3$ in general. The neutron diffraction 
pattern of TbFe$_{0.5}$Mn$_{0.5}$O$_3$ at 300~K is 
presented in Fig~\ref{fig_npd_300} as black plus signs. 
The nuclear structure at 300~K is refined 
in $Pbnm$ space group. In Fig~\ref{fig_npd_300}, 
the calculated pattern is shown as red solid line, 
difference curve as gray dotted line and the 
allowed Bragg peaks for
$Pbnm$ space group as vertical bars. From the 
magnetization measurements presented in 
Fig~\ref{fig_mag} (a, e), it is clear that
TbFe$_{0.5}$Mn$_{0.5}$O$_3$ undergoes a 
magnetic phase transition very close to 
300~K (notice from Fig~\ref{fig_mag} (e) that the
transition extends over a wide range from 280~K to 303~K). 
Hence, the diffraction data
at 300~K is refined with additional magnetic phase.
\begin{table}[!b]
\caption{\label{tab3} The possible magnetic structures of $R$FeO$_3$ allowed by symmetry. The space group is chosen as $Pbnm$  and $x$, $y$ and $z$ denote orientations parallel to the crystallographic directions $a$, $b$ and $c$. $R$ occupies $4c$ and Fe, $4b$ Wyckoff positions in this structure.}
\setlength{\tabcolsep}{11pt}
\begin{tabular}{llllll} \hline\hline
Irreps                            &   Space group                    &        $4b$                          &          $4c$            &     \\ \hline\hline
$\Gamma_1$                  & $Pbnm$                             &    $A_xG_yC_z$                &       $C_z$             &      \\  
$\Gamma_2$                 & $Pbn'm'$                            &    $F_xC_yG_z$                &       $F_xC_y$             &      \\ 
$\Gamma_3$                 & $Pb'nm'$                            &    $C_xF_yA_z$                &       $C_xF_y$             &      \\ 
$\Gamma_4$                 & $Pb'n'm$                            &    $G_xA_yF_z$                &       $F_z$             &      \\  \hline\hline
\end{tabular}   
\end{table}
In order to solve the magnetic structure at 300~K, 
the peaks below 2$\Theta \approx$ 40$^{\circ}$ 
were used to 
perform a $\bf k$-search to find the propagation 
vector. The utility called $\bf k$-search within 
FullProf suite of programs
was used for this purpose. Thus, $\bf k$=(000) 
was identified as the propagation vector. 
Representation analysis using $\bf k$(000) and 
$Pbnm$ nuclear 
cell lead to the listing of four possibilities -- 
$\Gamma_1$ ($Pbnm$), $\Gamma_2$ ($Pbn'm'$), 
$\Gamma_3$ ($Pb'nm'$) and $\Gamma_4$ ($Pb'n'm$) 
matching with the selection in Table~\ref{tab3}. 
They also match with the magnetic structures of 
orthoferrites already reported in the literature
\cite{deng2015magnetic}.
From the refinement trials it was noted that the 
representation $\Gamma_3$ contributes zero 
intensity to the Bragg peaks at 
(101) and (011) and hence can be excluded. 
A better visual fit to the experimental data 
and reasonable agreement factors were obtained 
for $\Gamma_4$
(the $R_\mathrm{mag}$ factors were, 
$\Gamma_1\approx$ 70; $\Gamma_2\approx$ 21; 
$\Gamma_3\approx$ 20 and $\Gamma_4\approx$ 5) 
and hence was accepted as the solution to 
the magnetic structure at 300~K. 
In Fig~\ref{fig_npd_300}, the lower set of 
vertical tick marks correspond to the magnetic Bragg positions.
\\
\indent
After confirming the room-temperature crystal 
structure to be $Pbnm$ and the magnetic structure 
as $\Gamma_4$ ($Pb'n'm$),
we now discuss the low temperature diffraction data.
In Fig~\ref{fig_npd}, an expanded view of the 
reflections in the 2$\Theta$-range 31--35$^{\circ}$ 
is given for 2~K,
90~K and 300~K. At 2~K, the intensity of the (011) 
reflection is observed to increase, compared to 
the value at 300~K. 
The relative intensity, $I_{(011)}/I_{(101)} > 1$ 
at 2~K and 300~K where as the opposite is true for 90~K. 
Macroscopic magnetic characterization of 
TbFe$_{0.5}$Mn$_{0.5}$O$_3$
\cite{hariharan2015reorientation} 
clearly suggested spin-reorientations and 
magnetic phase transformations as a function 
of temperature and magnetic fields. 
At 250~K, the nuclear structure was refined in 
$Pbnm$ and the magnetic structure in 
$\Gamma_4$ ($Pb'n'm$) similar
to the case of 300~K-data. However, at 200~K 
mixed magnetic domains consisting of ($\Gamma_1$ + $\Gamma_4$) 
is found to reproduce the experimental data faithfully. 
Neither of the representations $\Gamma_1$ or $\Gamma_4$ 
alone could give a satisfactory fit, while the mixed-domain
model significantly returned lower values of reliability 
factors ($R\sim$ 5.3 for mixed-domains
whereas $\sim$ 6 for pure phases). A better fit using 
the mixed-domain model could result from utilizing a bigger
parameter space for least-squares however, we also 
notice the presence of clear irreversibility in the 
magnetization profiles that are in support of the 
claim of presence of mixed-domains. 
The mixed-domains of ($\Gamma_1$ + $\Gamma_4$) 
were found to exist down till $T^\mathrm{Fe/Mn}_{SR}$ at 26~K. 
Fig~\ref{fig_mag} shows that "loop-like" anomalies are 
present in the $M(T)$ which correspond 
to the spin-reorientation transition at 
$T^\mathrm{Fe/Mn}_{SR}$. In fact, from the panel (c) 
of Fig~\ref{fig_mag} it can be seen that the 
irreversibility in ZFC and FC curves commences at 36~K itself. 
Finally at 10~K, the $\Gamma_4$ representation 
observed at 300~K re-emerges. It was observed that the 
inclusion of Tb moments in the refinement at 
10~K and 5~K did not lead to any appreciable 
values of refined magnetic moment on the Tb 
site. At 2~K, the magnetic structure remains 
in the $\Gamma_4$ representation however, Tb 
develops a magnetic moment
value of 0.6(2)~$\mu_\mathrm{B}/$f.u. indicating 
that Tb is magnetically ordered at this temperature.
The irreversibility in $M (T)$ suggesting magnetic
ordering in the rare earth sublattice occurs $\approx$
5~K. However, attempts to refine the 5~K-data
assuming magnetic contribution from Tb did not
yield better agreement factors. Hence magnetic
contributions from Mn/Fe were only considered.
The refinement of the magnetic structure of Tb
was performed in $F_z$ representation following
the symmetry analysis\cite{deng2015magnetic}.
In Fig~\ref{fig_npd_refine} (a)--(e), the 
refined diffraction patterns of 
TbFe$_{0.5}$Mn$_{0.5}$O$_3$ 
at 200, 50, 26, 10~K and 2~K are shown. Though 
no structural distortions are observed in the entire 
temperature range of the study, the magnetic 
structure is seen to evolve between the $\Gamma_4$ and
$\Gamma_1$ representation through mixed-domain regions. 
A schematic of the magnetic structures 
$\Gamma_1$ and $\Gamma_4$ are shown in 
Fig~\ref{fig_magstr}. In the case of 
$\Gamma_4$, the $F_z$ type ordering of 
Tb ions are also depicted (Tb moments
are shown in blue).
Next, we make an attempt to quantify the volume
fraction of the two magnetic domains $\Gamma_1$
and $\Gamma_4$. For this calculation we assume 
that the total Fe/Mn ordered moment at a particular 
temperature is equal for both $\Gamma_1$ and $\Gamma_4$ 
domains. With this assumption and by setting the 
scale factors of both $\Gamma_1$ and $\Gamma_4$ domains 
equal to that of the nuclear phase, the ratio of refined 
ordered moments of $\Gamma_1$ and $\Gamma_4$ should 
give the percentage of each domains. The result of
this calculation is presented as Fig~\ref{fig_domains}.
As clear from the figure, mixed domain structure is
present in the wide temperature range of 250~K to 26~K.
Outside this temperature window, the major magnetic
phase is that of $\Gamma_4$.
\section{\label{DISCUSSION}Discussion}
\indent
Our investigation of the magnetic structure 
of TbFe$_{0.5}$Mn$_{0.5}$O$_3$
shows that the predominant structures are 
$\Gamma_4$ and $\Gamma_1$
however, a competition between these two 
magnetic phases is evident from 
the presence of a mixed-domains in certain 
temperature range. 
At 295~K, the high temperature paramagnetic phase
transforms to the magnetic structure 
$\Gamma_4$ ($G_xA_yF_z$ or $Pb'n'm$). 
It is seen that down till $\approx$ 250~K, 
the $\Gamma_4$ structure remains stable. In the 
intermediate temperature range close to 200~K 
a mixed-domains of ($\Gamma_1$ + $\Gamma_4$) 
is observed which remains down till 26~K. 
A precise phase space study to determine 
the boundaries of the mixed phase would 
demand many more temperature points, 
but such a detailed study was outside the 
design and scope of our present experiments.
\begin{figure}[!t]
\centering
\includegraphics[ scale=0.32]{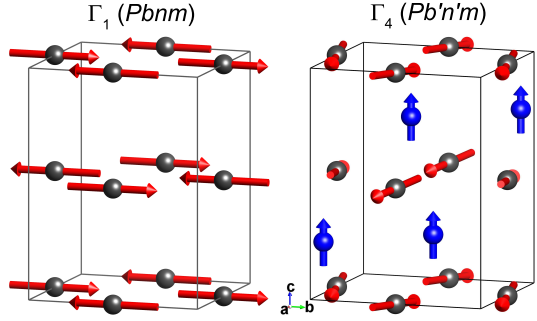}
\caption{\label{fig_magstr} (color online) The (left) $\Gamma_1$ and (right) $\Gamma_4$ magnetic structures of TbFe$_{0.5}$Mn$_{0.5}$O$_3$. For the $\Gamma_4$ structure, the Tb magnetic moments are also presented (in blue). The magnitude of Tb moments have been scaled $\times$2 for the sake of clarity in viewing.}
\end{figure}
As evidenced by the $M(T)$ curve 
(Fig~\ref{fig_mag} (a)), the $T^\mathrm{Fe/Mn}_{SR}$ 
is associated with a clear bifurcation of ZFC 
and FC forming a "loop-like" region between 18~K and 36~K. 
We deduce that the mixed-domains extend over this 
temperature window. Further at 10~K, the $\Gamma_4$ 
structure reemerges and remains till 2~K.
In the temperature range 2 - 10~K, the magnetic 
representation $\Gamma_1$ leads to a high 
$R_\mathrm{mag}\approx$ 60. It did not adequately account for 
the magnetic peaks and hence, $\Gamma_4$ 
($R_\mathrm{mag}\approx$ 3.5) was chosen as the solution. 
Note that in both $\Gamma_1$ and $\Gamma_4$, Fe/Mn 
are constrained to have magnetic moments in $x$, $y$ and $z$ 
whereas the $R$-moment is constrained to the $z$-direction. 
We found that in whole temperature range in $\Gamma_1$ 
setting the Fe/Mn magnetic moments had negligible $x$ and $z$ 
components (practically zero) and aligned along y-axis 
(Fig. 6). Also the whole temperature range in $\Gamma_4$ 
setting Fe/Mn moments have negligible y-component and aligned along 
$x$-axis with small canting along $y$-direction.
\\
\begin{figure}[!t]
\centering
\includegraphics[ scale=0.42]{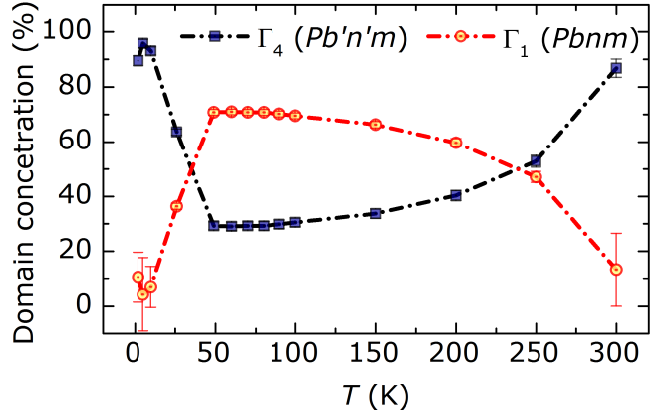}
\caption{\label{fig_domains} (color online) The volume fraction of the different magnetic domains $\Gamma_1$ and $\Gamma_4$ plotted as a function of temperature. Between 26~K and 250~K the co-existence of both the domain fractions lead to the irreversibility seen in magnetization measurements.}
\end{figure}
\indent
Different rare earths in $R$FeO$_3$ are observed to have 
different magnetic structures at low temperatures. For example, 
DyFeO$_3$ has a $Pb'n'm'$ space group for Dy and $Pnma$ for Fe whereas 
TbFeO$_3$ has $Pbnm'$ for Tb. In the present case of 
TbFe$_{0.5}$Mn$_{0.5}$O$_3$, it is found that in the low 
temperature region significant diffuse scattered intensity is present 
arising from the short-range magnetic order of Tb. 
Supporting this feature is the fact that no enhancement on quality 
of the fit was obtained in refining the 5~K or 10~K diffraction data by 
introducing a magnetic moment for Tb. Only at 2~K does Tb begin to order 
as $F_z$ as evident by the development of 0.6(2)~$\mu_\mathrm{B}$ for 
Tb moment. A notable difference of the magnetic structure of Fe/Mn in 
TbFe$_{0.5}$Mn$_{0.5}$O$_3$ compared to that of the parent compound TbFeO$_3$ 
is that the latter compound transforms to $\Gamma_2$ ($F_xC_yG_z$) structure 
at the spin-reorientation transition. In both the cases, the high 
temperature structure pertaining to Fe/Mn is $\Gamma_4$. 
In TbFeO$_3$, the Tb moment undergoes two types of ordering below 
10~K in to $F_xC_y$ and to $A_xG_y$ whereas in 
TbFe$_{0.5}$Mn$_{0.5}$O$_3$, Tb develops no significant 
magnetic moment until 2~K where it orders $F_z$.
\section{\label{CONCLUSION}Conclusions}
In conclusion, the Mn-doped orthoferrite compound TbFe$_{0.5}$Mn$_{0.5}$O$_3$ orders antiferromagnetically at $T^\mathrm{Fe/Mn}_{N}\approx$ 300~K and undergoes spin-reorientation transition at $T^\mathrm{Fe/Mn}_{SR}$ 26~K. Further, at 2~K the rare earth Tb is found to be magnetically ordered, however, the exact ordering temperature is no confirmed through our study. It is found that in the intermediate temperatures 250~K to 26~K, a mixed-domain model best describe the neutron diffraction data. At 2~K and at 300~K $\Gamma_4$ representation is stable while at the spin-reorientation transition and around 200~K mixed-domains of ($\Gamma_1$ + $\Gamma_4$) exist. An estimate of the domain concentration as a function of temperature is made. Clear indication of diffuse magnetic scattering from Tb is present especially below the $T^\mathrm{Fe/Mn}_{SR}$ while long-range order emerges at 2~K.
\section*{Acknowledgements}
H. S. N. acknowledges FRC/URC of UJ for a postdoctoral fellowship. A. M. S. thanks the SA-NRF (93549) and the FRC/URC of UJ for financial assistance. 
%
%
%
%
%
%
%
%
\end{document}